\documentclass[a4paper,fleqn,usenatbib]{mnras}

\usepackage{newtxtext,newtxmath}

\usepackage[T1]{fontenc}
\usepackage{ae,aecompl}
\usepackage[normalem]{ulem}

\usepackage{graphicx}	
\usepackage{amsmath}	

\usepackage[usenames,dvipsnames]{xcolor}

\newcommand{\fig}[1]{Figure~\ref{fig:#1}}
\renewcommand{\sec}[1]{Section~\ref{sec:#1}}

\newcommand{\App}[1]{Appendix~\ref{sec:#1}}
\newcommand{\Eq}[1]{Equation~\ref{eq:#1}}
\newcommand{\Fig}[1]{Figure~\ref{fig:#1}}

\newcommand{\lcdm}{\mbox{$\Lambda$CDM}}
\newcommand{\eagle}{\mbox{\sc{Eagle}}}
\newcommand{\flares}{\mbox{\sc{Flares}}}

\title[Extreme Value Statistics at High-$z$]{Extreme Value Statistics of the Halo and Stellar Mass Distributions at High Redshift: are JWST Results in Tension with $\Lambda$CDM?}

\author[C. C. Lovell et al.]{Christopher C. Lovell,$^{1,2}$\thanks{E-mail: christopher.lovell@port.ac.uk}
Ian Harrison$^{3}$,  \newauthor
Yuichi Harikane$^{4}$,
Sandro Tacchella$^{5,6}$, 
Stephen M. Wilkins$^{7}$
\\
$^{1}$Institute of Cosmology and Gravitation, University of Portsmouth, Burnaby Road, Portsmouth, PO1 3FX, UK\\
$^{2}$Centre for Astrophysics Research,  School of Physics, Engineering \& Computer Science, University of Hertfordshire, Hatfield AL10 9AB, UK\\
$^{3}$School of Physics and Astronomy, Cardiff University, The Parade, Cardiff, CF24 3AA, UK\\
$^{4}$Institute for Cosmic Ray Research, The University of Tokyo, 5-1-5 Kashiwanoha, Kashiwa, Chiba 277-8582, Japan\\
$^{5}$Kavli Institute for Cosmology, University of Cambridge, Madingley Road, Cambridge, CB3 0HA, UK\\
$^{6}$Cavendish Laboratory, University of Cambridge, 19 JJ Thomson Avenue, Cambridge, CB3 0HE, UK\\
$^{7}$Astronomy Centre, University of Sussex, Falmer, Brighton BN1 9QH, UK\\
}

\date{Accepted XXX. Received YYY; in original form ZZZ}

\pubyear{2022}

\begin{document}
\label{firstpage}
\pagerange{\pageref{firstpage}--\pageref{lastpage}}
\maketitle

\begin{abstract}
The distribution of dark matter halo masses can be accurately predicted in the \lcdm\ cosmology.
The presence of a single massive halo or galaxy at a particular redshift, assuming some baryon and stellar fraction for the latter, can therefore be used to test the underlying cosmological model.
A number of recent measurements of very large galaxy stellar masses at high redshift ($z > 8$) motivate an investigation into whether any of these objects are in tension with \lcdm.
We use extreme value statistics to generate confidence regions in the mass-redshift plane for the most extreme mass haloes and galaxies.
Tests against numerical models show no tension, neither in their dark matter halo masses nor their galaxy stellar masses.
However, we find tentative $> 3\sigma$ tension with recent observational determinations of galaxy masses at high redshift from both HST \& JWST, despite using conservative estimates for the stellar fraction ($f_{\star} \sim 1$).
Either these galaxies are in tension with \lcdm, or there are unaccounted for uncertainties in their stellar mass or redshift estimates.
\end{abstract}




\section{Introduction}

In the \lcdm\ paradigm, structure forms hierarchically in a bottom-up fashion, whereby density perturbations in the matter distribution at the time of inflation collapse first, then merge to form larger and larger structures.
Within this framework, baryons fall in to virialised dark matter haloes and form galaxies \citep{somerville_physical_2015}.
At late times ($z < 2$), the largest overdensities collapse to form galaxy clusters, $> 10^{14} \; \mathrm{M_{\odot}}$ haloes hosting a hot, x-ray emitting intracluster medium and hundreds, sometimes thousands of galaxies.
At earlier times ($z > 2$) clusters have yet to form; galaxies and their host haloes are the largest virialised objects in the universe.

In this standard `concordance' cosmology the predicted halo mass distribution is trivial to calculate.
It can then be used to constrain deviations from this concordance picture, for example the effect of non-gaussian initial conditions \citep[\textit{e.g.}][]{matarrese_abundance_2000,jimenez_implications_2009}.
One approach exploits Extreme Value Statistics \citep[EVS;][]{gumbel_statistics_1958, katz_extreme_2000}, which seeks to make predictions for the greatest (or least) valued random variable drawn from an underlying distribution.
The power of EVS is that it allows a test of the underlying cosmology from the observation of a \textit{single} extreme object.
It also provides both upper- and lower-limits on the mass of that object.
\cite{harrison_exact_2011} applied EVS to the predicted halo mass function to generate the Probability Density Function (PDF) of the most massive halo at a given redshift.
They extended this in \cite{harrison_testing_2012} to survey volumes in order to assess whether any observed high redshift clusters ($1 < z < 2$) exceeded the maximum expected mass according to $\Lambda$CDM, finding no tension between observations and theory \citep[see also][]{waizmann_application_2012,chongchitnan_primordial_2012}.
The approach has also been applied to the distribution of void sizes \citep{chongchitnan_abundance_2015,sahlen_clustervoid_2016}.

In order to extend this approach to higher redshifts we require measurements of much lower halo masses than those hosting galaxy clusters.
Unfortunately, such measurements are difficult, particularly at high redshift.
Halo masses can be inferred from galaxy clustering, which has the benefit of not needing to assume the underlying baryonic physics, but cannot be used to measure the masses of individual objects.
Abundance matching fixes the knee of the halo mass function to the knee of an observed luminosity function, but this explicitly uses features of the dark matter model to infer the halo masses.
Another method is to measure the direct emission from the baryonic components of a galaxy, and assume some scaling with the total mass, or use Spectral Energy Distribution (SED) modelling to estimate the baryonic masses.
These masses can then be combined with the cosmological baryon fraction, $f_{\mathrm{b}}$, and subsequent fractions of the relevant components, \textit{e.g.} the stellar mass fraction $f_{\star}$, to derive the latent halo mass.

\cite{steinhardt_impossibly_2016} first explored this approach, assuming a mass--to--light ratio measured at $z = 4$ and a fixed stellar--halo mass relation, finding some tension with observations at $4 < z < 8$.
\cite{behroozi_most_2018} also implement this latter method, using the cosmic baryon fraction as an absolute upper limit on the ratio of galaxy stellar mass to halo mass ($f_{b} \sim 0.16$; \citealt{planck_collaboration_planck_2016}), whilst also allowing for redshift evolution in the stellar--halo mass relation.
This relationship can be inverted to place an upper limit on the halo mass for an observed stellar mass, and then compared with predicted halo mass functions.
They found that, at the time, no observed galaxies exceeded these conservative upper limits.
More recently, \cite{boylan-kolchin_stress_2022} used a similar approach to \cite{behroozi_most_2018} to test whether any recent high-mass, high-$z$ candidates discovered in the first JWST data \citep{donnan_evolution_2022, harikane_comprehensive_2022, labbe_very_2022, naidu_two_2022, finkelstein_long_2022, adams_discovery_2022, rodighiero_jwst_2022} exceed the limits set by \lcdm.
Uniquely, they test both the number density of galaxies above some stellar mass at early epochs, as well as limits placed on the stellar mass \textit{density}.
They find strong tension, particularly with the latter, for the candidates presented by \cite{labbe_very_2022} at $z \sim 10$, but less tension with other studies.
\cite{menci_high-redshift_2022} have also used the abundance of high redshift JWST candidates to place constraints on dark energy models.

The EVS approach has a number of advantages over previous approaches.
EVS mitigates the problem of the selection function in galaxy surveys; the most massive object acts as a lower limit on the most massive object one could have seen in a given survey.
For estimates using full samples, uncertainty in the selection function can propagate into population measurements, \textit{e.g.} the mean mass of your sample.
As mentioned above, EVS also provides two--sided constraints (upper and lower limits) on the most massive object, and naturally considers the uncertainty in the mass of that object for a given survey volume / area.
By combining with realistic functional forms for the stellar and baryon fraction, EVS naturally incorporates uncertainty in these parameters.

In this paper we use EVS to calculate the full PDF of the mass of the most massive halo.
We first compare to numerical simulations, computing the EVS PDF on fixed redshift hypersurfaces and compare to individual snapshots taken from these simulations (\sec{hypersurface}).
We then proceed to calculate the statistics for observational survey volumes, and compare to recent observational measurements of galaxy masses, as well as make predictions for upcoming wide field surveys with JWST, Roman and Euclid (\sec{obs}).
Observations of galaxy or halo masses significantly greater than the expected values for the \emph{most massive} object would imply tension with $\Lambda$CDM.
Equally, by computing the full PDF with EVS, we can evalute the \textit{minimum} mass of the most massive halo or galaxy; if the largest observed object has a mass significantly lower than that predicted by EVS, this will place equally high tension on $\Lambda$CDM.
We discuss our results and present our conclusions in \sec{conc}.
We also present a python package for computing confidence intervals for arbitrary survey areas (\href{https://github.com/christopherlovell/evstats}{github.com/christopherlovell/evstats}).
We assume a flat $\Lambda$CDM cosmology with \cite{planck_collaboration_planck_2016} parameters: $\Omega_{\mathrm{M}} = 0.309$, $\Omega_{b} = 0.0486$, $\sigma_{8} = 0.816$, $h = 0.678$.

\section{Extreme Value Statistics on a Fixed Redshift Hypersurface}
\label{sec:hypersurface}

Extreme Value Statistics \citep[EVS;][]{gumbel_statistics_1958, katz_extreme_2000} is concerned with the most extreme deviations from the median of a probability distribution.
Consider a sequence of $N$ random variates $\{ M_{i} \}$ drawn from a cumulative distribution function (CDF), $F(m)$.
There will be a largest value of the sequence, $M_{\mathrm{max}} \equiv \mathrm{sup} \{ M_{1} ... M_{N}\}$.
Assuming all variables are mutually independent and identically distributed (IID), the probability all deviates are less than or equal to some value $m$ is given by:
\begin{align}
\Phi (M_{\mathrm{max}} \leqslant m;\, N) &= F_{1} (M_{1} \leqslant  m) \; ... \; F_{N} (M_{N} \leqslant  m) \\
&= F^{N} (M)
\label{eq:cdf}
\end{align}
By differentiating \Eq{cdf} we find the probability density function (PDF) of the distribution,
\begin{align}
\Phi (M_{\mathrm{max}} = m;\, N) &= N F'(m) [ F(m) ]^{N-1} \\
&= N f(m) [ F(m) ]^{N-1} \;,
\label{eq:pdf}
\end{align}
where $f(m)$ is the PDF of the original distribution ($f(m) = dF(m)/dm$), and
$\Phi (M_{\mathrm{max}} = m;\, N)$ is the \textit{exact} extreme value PDF for $N$ observations drawn from the known distribution $F(m)$.\footnote{For more details on the advantages of using the exact EVS statistics rather than those employing asymptotic theory, see \cite{harrison_testing_2012}}
We apply this general result to the case of the halo mass function, where $n(M)$ is the number density of haloes of mass $M$, and derive $f(M)$ and $F(M)$,
\begin{align}
f(m) &= \frac{1}{n_{\mathrm{tot}}} \frac{dn(m)}{dm} \;\;, \\
F(m) &= \frac{1}{n_{\mathrm{tot}}} \int^{m}_{- \infty} dM \frac{dn(M)}{dM} \;,
\end{align}
where $n_{\mathrm{tot}}$ is a normalisation factor giving the total (co-moving) number density of haloes,
\begin{equation}
n_{\mathrm{tot}} = \int^{\infty}_{- \infty} dM \frac{dn(M)}{dM} \;.
\end{equation}
Together, these equations can be used to estimate the EVS PDF for a constant redshift co-moving volume $V$, where the total number of haloes $N = n_{\mathrm{tot}} V$.\footnote{It is impractical and unnecessary to integrate between inifinite endpoints. We use conservative finite limits of $10^{6} \leqslant m \leqslant 10^{17}$ at all redshifts; the choice of these makes no difference to our results.}

The IID assumption will be broken where haloes are significantly clustered.
Where the volume probed is sufficiently large the distribution is essentially homogeneous.
A number of studies have shown that this volume limit is achieved above $\sim (100 \; \mathrm{Mpc})^3$ \citep{gelb_cold_1994,power_impact_2006,reed_halo_2007,davis_most_2011}.
Another consideration when comparing to periodic simulations is the impact of finite-volume effects on the abundance of galaxies.
In a given periodic volume there is a maximum fundamental mode that can be represented, and large scale power on scales greater than the size of the simulation volume will not be captured.
Additionally, only discrete modes can be represented in periodic volumes.
These approximations can impact the halo mass function, particularly at the high mass end where the effect of these large modes is more pronounced \citep{reed_halo_2007,derose_aemulus_2019}.
However, the volume at which these effects become pronounced has been shown to be $< (100 \; \mathrm{Mpc})^3$; \cite{lukic_halo_2007} show that the effect in boxes of this volume is < 10\% on the normalisation of the halo mass function (< 0.05 dex), subdominant to statistical error.
In this study we only analyse simulated and observational volumes above this limit.


\subsection{Halo Masses}

\begin{figure}
	\includegraphics[width=\columnwidth]{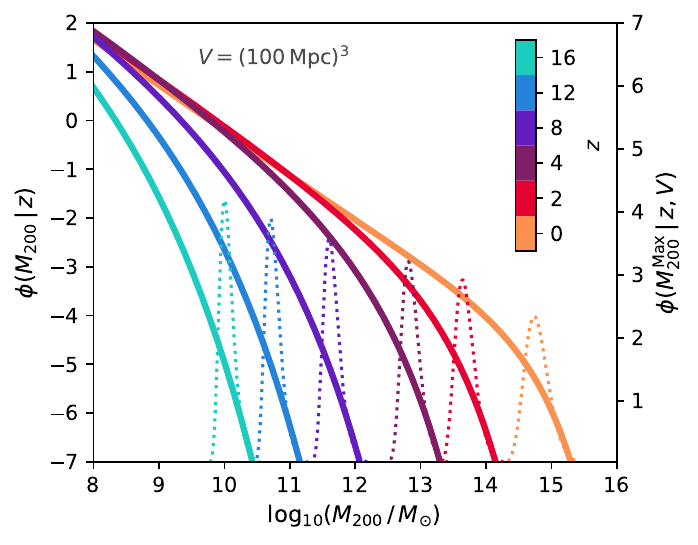}
    \caption{Halo mass function (solid lines; \protect\citealt{behroozi_average_2013}) in the range $0 \leqslant z \leqslant 20$, in units $\mathrm{Mpc}^{-3} \; \mathrm{dex}^{-1}$. The PDF, calculated using EVS, for the highest mass halo on a fixed hypersurface with volume $(100 \; \mathrm{Mpc})^{3}$ at each redshift is shown as a dotted line.}
    \label{fig:hmf}
\end{figure}

To calculate the EVS PDF we must first assume a form for the halo mass function.
We use the \cite{behroozi_average_2013} halo mass functions, which are calibrated using $N$-body simulations to the redshift range $2 \leqslant z \leqslant 8$ based on the \cite{tinker_toward_2008} mass functions (themselves derived from $0 < z \leqslant 2.5$ data).
The redshift evolution of the halo mass function parameters from \cite{behroozi_average_2013} has smoothed off at $z > 8$, so we assume it is safe to extrapolate to these redshifts; we also note that we have tested using other forms for the mass function, and found little impact on our conclusions.
\Fig{hmf} shows the halo mass function for a range of redshifts, along with the PDF for the highest mass halo on a constant redshift hypersurface, with volume (100 Mpc)$^3$, predicted by EVS.
The peak of the PDF corresponds to the most probable mass of the most massive halo in the volume at that redshift.

A number of comparisons of the predictions of EVS with numerical simulations have been carried out in the past \citep{harrison_testing_2012,watson_statistics_2014}, all showing consistency.
To test our results we compare to predictions for the most massive halo from two hydrodynamic cosmological simulations.
The fiducial EAGLE simulation \citep{schaye_eagle_2015,crain_eagle_2015} is a $(100 \; \mathrm{Mpc})^3$ cosmological volume evolving both dark matter and baryons self consistently.
The \flares\ simulations \citep{lovell_first_2021,vijayan_first_2021} use the EAGLE physics model to resimulate zooms of a range of overdensities during the epoch of reionisation, extending the dynamic range over periodic cosmological volumes.
Since \flares\ is not a continuous periodic box, one must calculate the `effective volume' of the combined zoom regions, which is dependent on the mass / luminosity of the selected galaxies.
We use a fixed effective volume of $(550 \; \mathrm{Mpc})^3$, which roughly corresponds to that for the most massive halo / galaxy at all redshifts.
The underlying halo mass function in both of these simulations is not identical to that presented by \cite{behroozi_average_2013}, but is in reasonably good agreement at the redshifts shown.

\begin{figure}
	\includegraphics[width=\columnwidth]{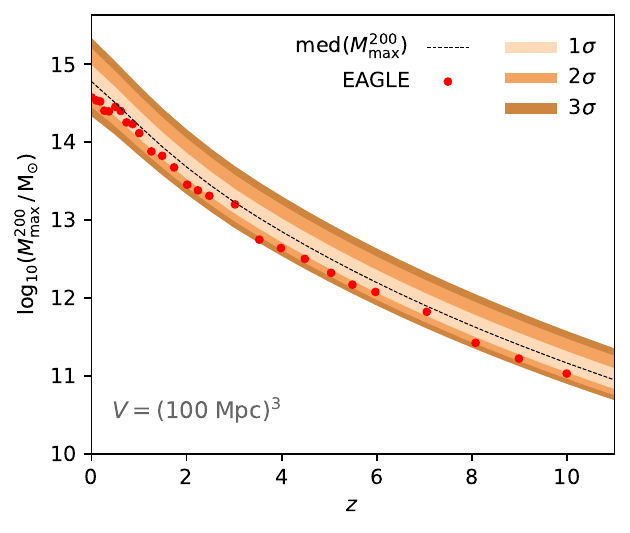}
    \includegraphics[width=\columnwidth]{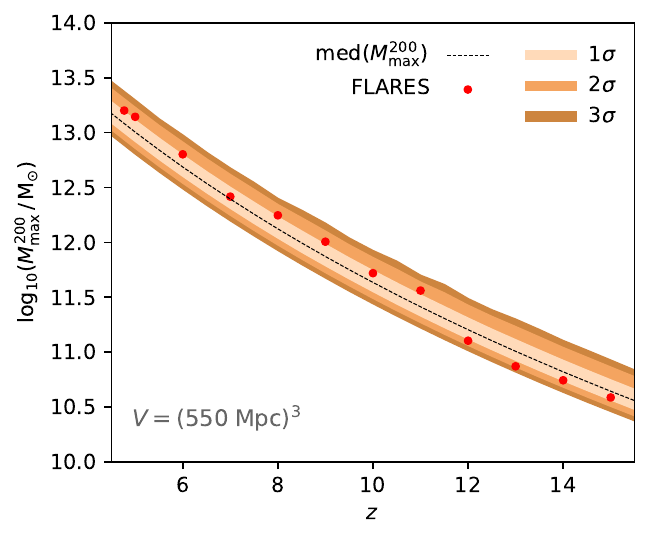}
    \caption{\textit{Top:} confidence intervals for the EVS PDF on a fixed redshift hypersurface, evaluated at a range of redshifts, for haloes taken from a $(100 \; \mathrm{Mpc})^3$ volume.
		The most massive halo from each available snapshot in the fiducial \eagle\ simulation (with identical volume) is shown in red.
        \textit{Bottom:} as above, but for the $(550 \; \mathrm{Mpc})^3$ effective volume of \flares, and showing the most massive halo at each available snapshot.
		}
    \label{fig:hypersurface_halo_EAGLE}
\end{figure}

In this example, as well as in the sections below, we wish to show the EVS PDF for a range of redshifts simultaneously.
To do this we calculate the PDF for narrow redshift intervals ($\Delta z = 0.2$, where the number of bins is chosen so that $N_{\mathrm{bins}} >> N_{\mathrm{galaxies}}$),\footnote{The choice of $\Delta z$ has negligible impact on the results.
For a single bin, the maximum mass in this bin is the same as that measured over multiple bins; due to hierarchical and positive structure formation, this tends to be biased towards those objects at lower redshifts.}
Each bin can be thought of as a Bernoulli trial, therefore there is a non-zero probability of exceeding a given contour threshold; we have tested and found that, for our chosen binning, this probability is negligible. Further discussion on this effect is provided in \App{z_bin}.
We integrate over these PDFs to find the [1,2,3]$\sigma$ confidence intervals, and plot these along with the median of each distribution.
\fig{hypersurface_halo_EAGLE} shows the PDF of $M_{200}$ evaluated at a range of redshifts, along with the value of $M_{200}$ from the most massive halo selected from each available simulation snapshot in \eagle\ and \flares.
All of the simulated halos lie within the reasonably tight $3 \sigma$ confidence intervals.
The level of agreement is very good, and gives us confidence that our EVS scheme is correctly able to produce realistic contours in the halo mass--redshift plane, despite assuming a slightly different halo mass function to that produced in the simulations.
We now introduce astrophysical effects to predict the stellar mass distribution.


\subsection{Stellar Masses}

\begin{figure}
    \includegraphics[width=\columnwidth]{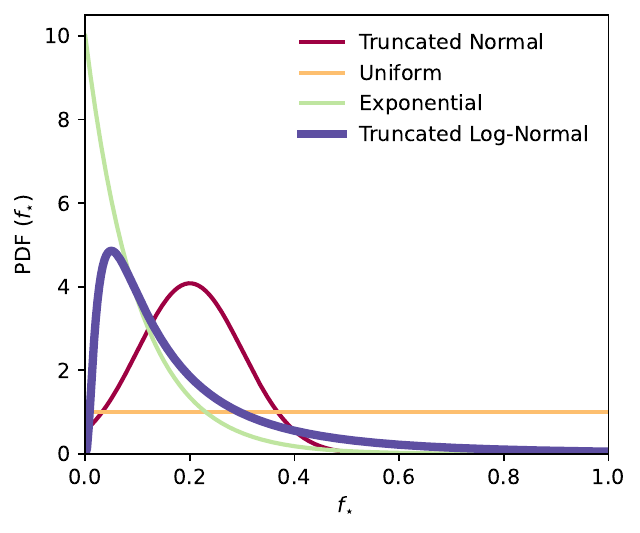}
	\includegraphics[width=\columnwidth]{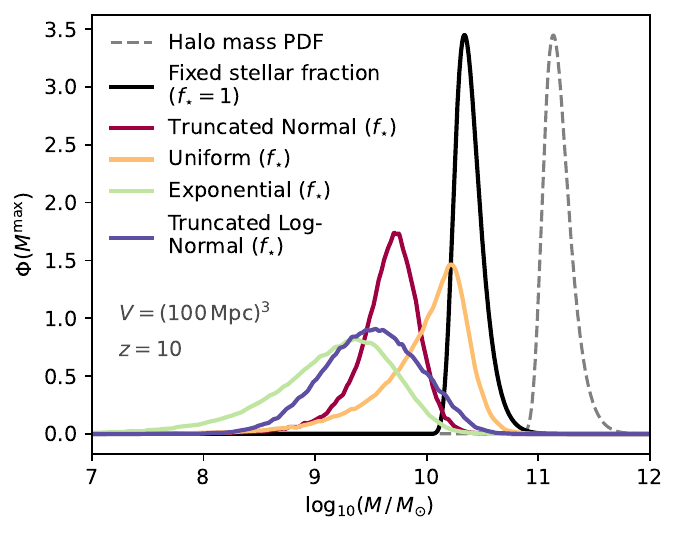}
    \caption{\textit{Top:} Parametric forms for the distribution of $f_{\star}$.
    \textit{Bottom:} stellar mass EVS PDF as a function of mass for a fixed redshift hypersurface at $z = 10$, assuming different parametric forms for $f_{\star}$.
		The halo mass PDF is shown by the grey dashed line.
		The black solid line shows the stellar mass PDF assuming $f_{\star} = 1$.
		}
    \label{fig:hypersurface_pdf_comparison}
\end{figure}

To convert our PDF for the halo distribution to a PDF for the galaxy stellar mass we must account for both the baryon fraction, $f_{\mathrm{b}}$, and the stellar fraction, $f_{\star}$.
The baryon fraction is set by our assumed cosmology \citep[$f_{\mathrm{b}} = 0.16$;][]{planck_collaboration_planck_2016}.
We assume a fixed value, though we note that this value can vary in different cosmic environments \citep[\textit{e.g.} lower than the universal value in local galaxy clusters][]{gonzalez_galaxy_2013}.
This is an effect that could be accounted for by using a functional distribution for $f_{\mathrm{b}}$, though we note that at high redshifts deviations from the universal value are not expected to be as large, due to the shorter time for feedback effects to have imprinted on baryon distributions.

The stellar fraction is dependent on the astrophysics that converts cold gas into stars.
A conservative upper limit is to assume \textit{all} baryons are converted into stars, $f_{\star} = 1$, and simply multiply the halo PDF by the product of the baryon and stellar fractions,
\begin{equation}
\Phi(M_{\mathrm{*}}) = \Phi(M_{\mathrm{DM}}) \, f_{\mathrm{b}} \, f_{\star} \;.
\end{equation}
The bottom panel of \fig{hypersurface_pdf_comparison} shows an example of the halo EVS PDF, as well as the stellar mass PDF obtained using a fixed stellar fraction of unity.
In reality, measurements of the stellar fraction suggest much lower values, particularly in the most massive halos \citep[\textit{e.g.}][]{giodini_stellar_2009}.
To account for this, we assume a (truncated; $0 \leqslant f_{\star} \leqslant 1$) lognormal distribution of $f_{\star}$,
\begin{align}
	f_{\star} = \mathrm{ln} \; N(\mu, \sigma^2)
\end{align}
where $\mu = e^{-2}$ and $\sigma = 1$.
This simple model ignores the dependence of the stellar fraction on redshift and halo mass, but incorportates the range of values inferred from simple halo models \citep{tacchella_physical_2013,tacchella_redshift-independent_2018} and observations \citep[\textit{e.g.}][]{harikane_evolution_2016,harikane_goldrush_2018,stefanon_galaxy_2021}.
However, it is worth noting that in the pre-reionization epoch ($z > 10$) high star formation efficiencies, close to the cosmic baryon fraction, have been predicted from theoretical models \citep{susa_effects_2004}.
We then calculate the product of this PDF with the halo PDF, whilst assuming the same fixed baryon fraction.
\fig{hypersurface_pdf_comparison} shows the log-normal form of $f_{\star}$ as well as other parametric forms, and an example of the stellar mass PDF obtained using these different distributions.
We also present upper $3\sigma$ limits based on assuming $f_{s} = 1$ and applying directly to the halo EVS PDF, which can be interpreted as conservative upper limits.

\begin{figure}
	\includegraphics[width=\columnwidth]{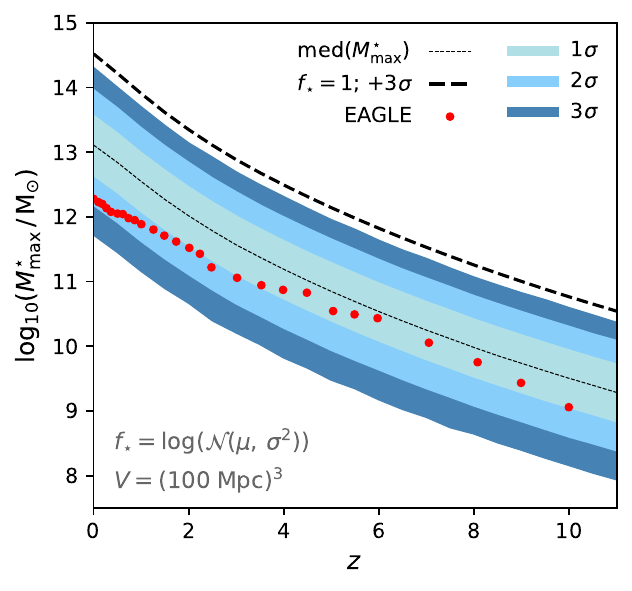}
    \includegraphics[width=\columnwidth]{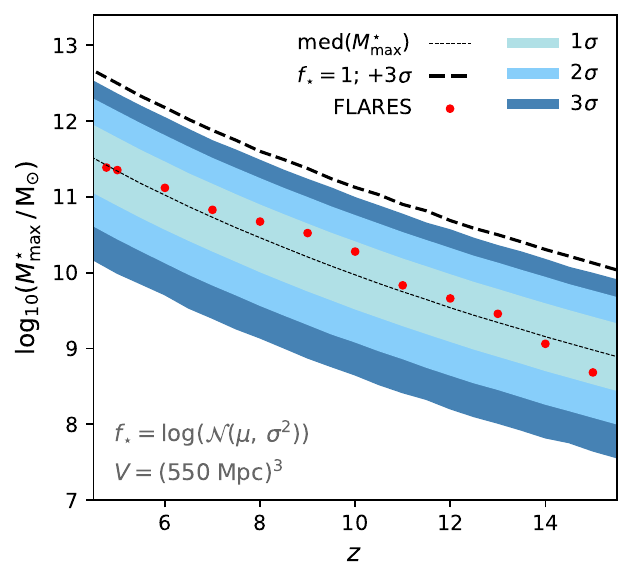}
    \caption{
        \textit{Top:} the stellar mass EVS confidence intervals on a fixed redshift hypersurface, evaluated at a range of redshifts, for galaxies taken from a $(100 \; \mathrm{Mpc})^3$ volume.
        The dashed line shows the $3\sigma$ upper limit assuming a stellar fraction of unity.
    	The most massive galaxy from each available snapshot in the fiducial EAGLE simulation (with identical volume) is shown in red.
        \textit{Bottom:} as above, but for the $(550 \; \mathrm{Mpc})^3$ effective volume of \flares, showing the most massive galaxy at each available snapshot from all resimulations.
    }
    \label{fig:hypersurface_mstar_EAGLE}
\end{figure}

\fig{hypersurface_mstar_EAGLE} shows the stellar mass PDF for a fixed redshift hypersurface, with volume $(100 \; \mathrm{Mpc})^3$ and $(550 \; \mathrm{Mpc})^3$, assuming this lognormal distribution of $f_{\star}$.
The uncertainties are larger than for the halo PDF as expected.
Results from the \eagle\ and \flares\ simulation are shown, and all lie within the uncertainties.
There is a noticeable plateau in the maximum stellar mass in EAGLE as we go to lower redshifts, which demonstrates the redshift and halo-mass dependent evolution of $f_{\star}$, particularly at low-$z$.
For now we ignore this redshift and halo-mass dependence, and assume a fixed distribution, however one could incorporate these effects.


\section{Extreme Value Statistics for an Observational Survey Volume}
\label{sec:obs}

In a galaxy survey we do not observe galaxies at a fixed redshift, and must therefore take account of the change in volume with redshift in an expanding universe, as well as the change in the number density of haloes with redshift due to the growth of structure.
The PDF and CDF for haloes in a fixed fraction of the sky, $f_{\mathrm{sky}}$, between redshifts $z_{\mathrm{min}}$ and $z_{\mathrm{max}}$ is then given by:
\begin{align}
f(m) &= \frac{f_{\mathrm{sky}}}{n_{\mathrm{tot}}} \left [ \int_{z_{\mathrm{min}}}^{z_{\mathrm{max}}} dz \frac{dV}{dz} \frac{dn(m,z)}{dm} \right ] \\
F(m) &= \frac{f_{\mathrm{sky}}}{n_{\mathrm{tot}}} \left [ \int_{z_{\mathrm{min}}}^{z_{\mathrm{max}}} \int^{m}_{- \infty}  dz \, dM \frac{dV}{dz} \frac{dn(M,z)}{dM} \right ] \;,
\end{align}
where
\begin{equation}
n_{\mathrm{tot}} = f_{\mathrm{sky}} \left [ \int_{z_{\mathrm{min}}}^{z_{\mathrm{max}}} \int^{\infty}_{- \infty}  dz \, dM \frac{dV}{dz} \frac{dn(M,z)}{dM} \right ] \;.
\end{equation}
We can then use these with \Eq{pdf} to find the halo EVS for a given survey.
As in \sec{hypersurface}, we assume the \cite{behroozi_average_2013} halo mass functions, a fixed baryon fraction, $f_{\mathrm{b}} = 0.16$, and a truncated lognormal distribution for the stellar fraction.

\subsection{Eddington Bias}
To compare our theoretical mass functions with observations we need to correct for Eddington Bias \citep{eddington_formula_1913}.
For haloes, the steepness of the mass function means there are significantly more low mass haloes than high mass, so there is greater upscatter of low mass halo measurements than downscatter of higher mass halo measurements, boosting the apparent number of higher mass haloes.
The same effect applies to galaxy stellar masses.
We correct using the following,
\begin{equation}
\mathrm{ln} \; M_{\mathrm{edd}} = \mathrm{ln} \; M_{\mathrm{obs}} + \frac{1}{2} \, \epsilon \, \sigma^{\, 2}_{\, \mathrm{ln} \, M} \,\,,
\end{equation}
where $\epsilon$ is the local slope of the underlying halo mass function, and $\sigma_{\, \mathrm{ln} \, M}$ is the uncertainty in the halo / stellar mass estimate.
We choose to correct the observations, using quoted uncertainties on the stellar mass.
For stellar masses we derive $\epsilon$ from the halo mass function, but use the halo mass given by the observed stellar mass multiplied by the inverse baryon fraction.
The true steepness of the galaxy stellar mass function is known to be steeper at low-$z$ due to AGN feedback, but this effect is expected to be less extreme at the high redshifts considered here.

\begin{figure}
	\includegraphics[width=\columnwidth]{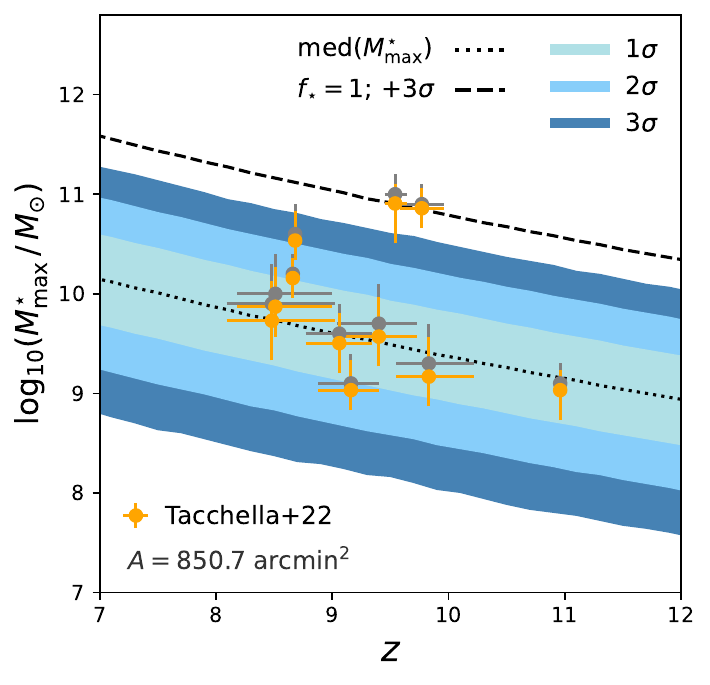}
    \includegraphics[width=\columnwidth]{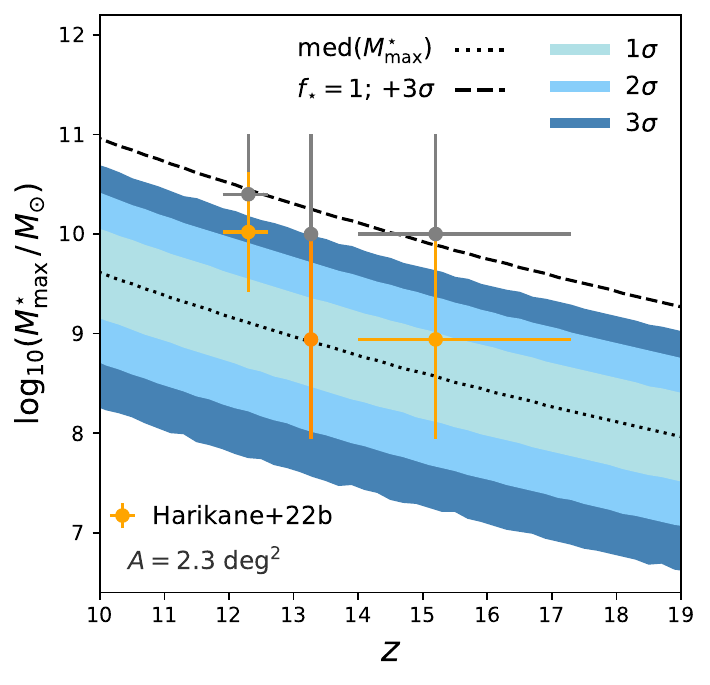}
    \caption{
        \textit{Top:} The stellar mass EVS confidence intervals for an observational survey volume with area $850.7 \; \mathrm{arcmin}^2$, evaluated at a range of redshifts.
		Stellar mass estimates from \protect\cite{tacchella_stellar_2022} for galaxies selected from the \textit{HST}/CANDELS fields \protect\citep{finkelstein_census_2022}, are shown in yellow, after correcting for Eddington bias.
        Grey points show the uncorrected stellar mass estimates.
		The dashed line shows the $3\sigma$ upper limit assuming a stellar fraction of unity.
        \textit{Bottom:} as above, but showing an observational survey volume with area $2.3 \; \mathrm{deg}^2$, and observational results from \protect\cite{harikane_search_2022} converted to a stellar mass estimate using a linear scaling relation.
		}
    \label{fig:evs_finkelstein21}
\end{figure}

\subsection{Observational Comparison (pre-JWST)}

A number of recent studies have presented estimates of galaxy stellar masses at high redshift ($z > 8$).
\cite{tacchella_stellar_2022} present an anlysis of a number of bright galaxies selected from \textit{HST} CANDELS fields, with associated \textit{Spitzer}/IRAC fluxes \cite{finkelstein_census_2022}.
They use the \textsc{prospector} \citep{leja_deriving_2017,johnson_stellar_2021} Bayesian SED fitting code to obtain stellar mass estimates from the photometric data, then present an analysis of the likelihood of the stellar mass estimates in \lcdm\ using the methodology of \cite{behroozi_most_2018}.
They probe down to some limiting number density $\Phi > 10^{-6} \; \mathrm{Mpc^{-3}}$, approximately that expected for a similar survey area, and assume a baryon fraction $f_{\mathrm{b}} = 0.16$ and a conservative stellar fraction of unity, $f_{\star} = 1$.
Two galaxies in their sample are in tension with these constraints, COSMOS-20646 and UDS-18697, at $3\sigma$ and $4.6\sigma$, respectively.
However, they argue that cosmic variance \citep{trenti_cosmic_2008}, observational uncertainties (particularly the contribution of near neighbours contaminating the IRAC photometry), and measurement uncertainty (related to SED modelling assumptions, such as the choice of prior on the star formation history) reduce this tension significantly.

Here, we repeat their analysis using the EVS framework.
The top panel of \Fig{evs_finkelstein21} shows the galaxy stellar mass EVS PDF for the combined survey area of \cite{finkelstein_case_2015}, with the stellar mass estimates for the selected galaxies from \cite{tacchella_stellar_2022} after correcting for Eddington bias.
The majority of the galaxies are within the $3\sigma$ uncertainties, however the same two galaxies identified in \cite{tacchella_stellar_2022} lie outside these bounds, even after correcting for Eddington bias.
They lie on the $3\sigma$ upper limit assuming a stellar fraction of unity.

Recently, \cite{harikane_search_2022} presented two bright galaxy candidates, HD1 and HD2, from Hyper Suprime-Cam, VISTA, and \emph{Spitzer} observations of the COSMOS and SXDS fields.
Their photometry suggests redshifts of $z = 15.2^{+1.2}_{-2.1}$ and $12.3^{+0.4}_{-0.3}$, and one of the sources (HD1) additionally has a tentative detection of [O\textsc{iii}]$88\mathrm{\mu m}$, giving a spectroscopic redshift of $z=13.27$.
Estimates of physical properties for these sources are not well constrained, however they quote stellar masses in the range $10^{9}-10^{11}$ and $10^{9.8}-10^{11} \, M_{\odot}$ for HD1 and HD2, respectively.
These ranges bound the stellar masses obtained using the $M_{\star} - M_{\mathrm{UV}}$ relation at $z = 8$ from \cite{song_evolution_2016}.

The bottom panel of \fig{evs_finkelstein21} shows the galaxy stellar mass EVS PDF for the combined survey area ($2.3 \; \mathrm{deg}^2$), as well as the stellar mass estimates for HD1 and HD2 after correcting for Eddington bias.
We show HD1 using both the photometric and spectroscopic redshift estimates.
The stellar mass correction due to Eddington bias is quite large due to the significant uncertainties in the stellar masses, which brings the estimates within the $3\sigma$ contours for both objects.

\begin{figure*}
    \includegraphics[width=\columnwidth]{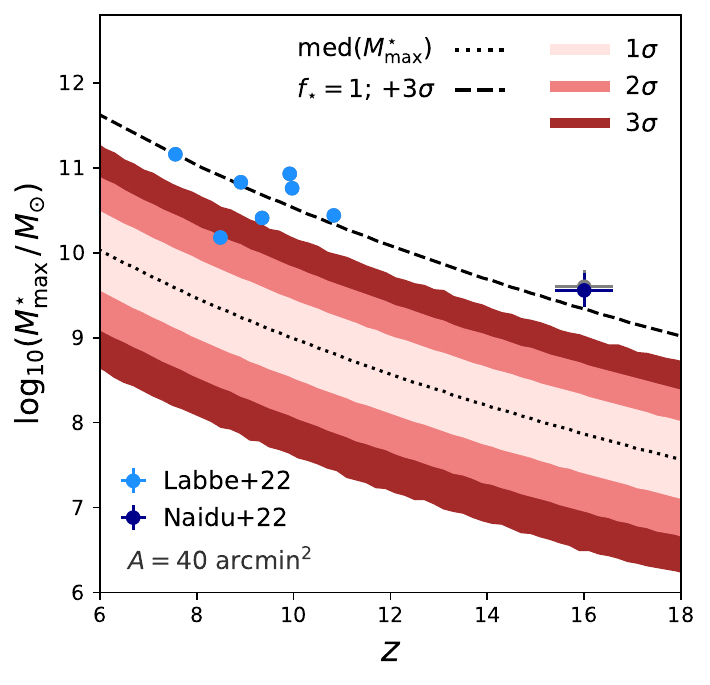}
    \includegraphics[width=\columnwidth]{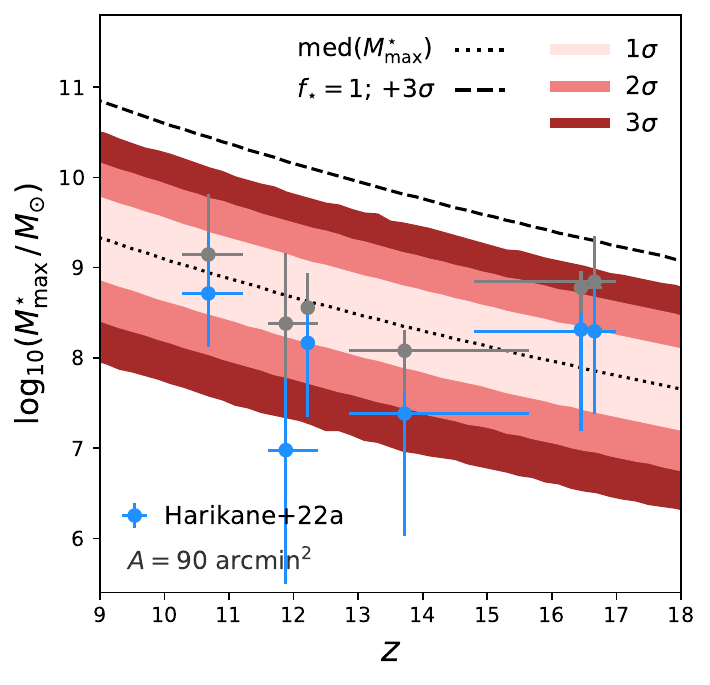}
    \includegraphics[width=\columnwidth]{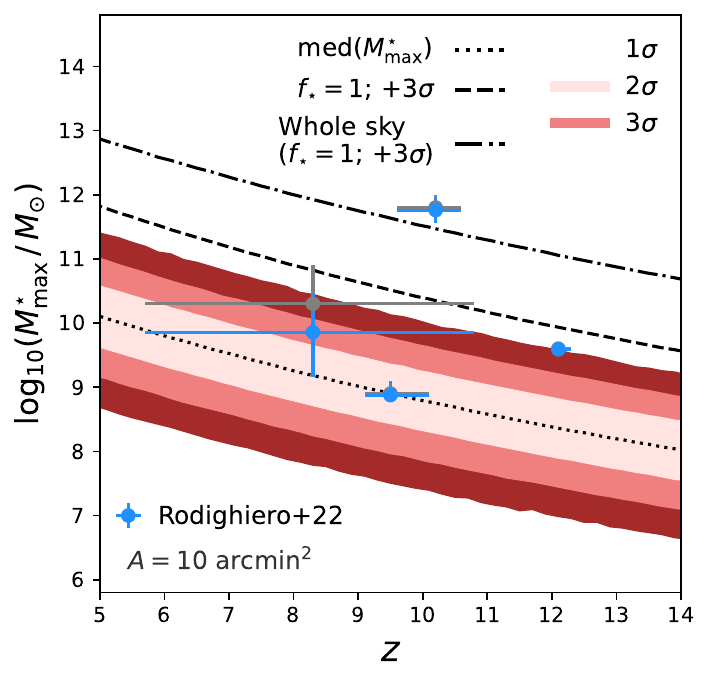}
    \caption{
        As for \fig{evs_finkelstein21}, but showing the latest high-$z$ candidates from JWST.
        \textit{Top left:} an observational survey volume with area $40 \; \mathrm{arcmin}^2$. Observational results from \protect\cite{labbe_very_2022} at $z \sim 10$ are shown, as well as the \protect\cite{donnan_evolution_2022} candidate, with stellar mass and redshift solutions at $z \sim 17$ derived by \protect\cite{naidu_schrodingers_2022} shown.
        \textit{Top right:} an observational survey volume with area $90 \; \mathrm{arcmin}^2$, with stellar mass estimates from \protect\cite{harikane_comprehensive_2022}.
        \textit{Bottom:} stellar mass estimates from \protect\cite{rodighiero_jwst_2022}, assuming an observational survey with area $10 \; \mathrm{arcmin}^2$. The dashed-dotted black line shows the $3\sigma$ upper limit, assuming a stellar fraction of unity, for a \textit{whole sky} survey. One of the \protect\cite{rodighiero_jwst_2022} candidates exceeds even this most conservative upper limit.
		}
    \label{fig:evs_jwst}
\end{figure*}

\subsection{Observational Comparison (JWST first results)}
\label{sec:obs_jwst}

In the short time since the first data from JWST was released there have been a number of studies estimating the redshifts and stellar masses of high redshift galaxies \citep[\textit{e.g.}][]{adams_discovery_2022,donnan_evolution_2022, finkelstein_long_2022, harikane_comprehensive_2022, labbe_very_2022, naidu_two_2022, rodighiero_jwst_2022}.
Many of these have proposed candidates that lie at the extremes of the redshift--stellar mass plane.
Here we test using the EVS framework whether any of these candidates are in tension with \lcdm.

\cite{labbe_very_2022} presented seven $> 10^{10} \, \mathrm{M_{\odot}}$ candidates at $7 < z < 11$ in a 40 arcmin$^2$ area, taken from the CEERS program, using \textsc{Eazy} and \textsc{Prospector} for the photometric redshift and stellar mass estimates, respectively.
The top left panel of \fig{evs_jwst} shows these candidates on the stellar mass--redshift plane, with the EVS PDF for an identical survey area.
5 out of the 7 candidates lie above the $3\sigma$ upper limits assuming a stellar fraction of unity.
Since they do not provide error estimates on the stellar masses we cannot evaluate the effect of Eddington bias, but if we assume 0.3 dex errors this relieves the tension with 3 of these candidates, leaving only 2 outside of the $3\sigma$ contours assuming a lognormal stellar fraction.

A particularly exciting discovery in the early data is a potential $z \sim 17$ candidate, also identified in the CEERS data (40 arcmin$^2$), first presented by \cite{donnan_evolution_2022}.
There is some debate as to the accuracy of this photometric redshift estimate \citep[\textit{e.g.}][]{zavala_dusty_2022}, with a $z \sim 5$ solution potentially also capable of explaining the observed photometry.
\cite{naidu_schrodingers_2022} provide stellar mass and photometric redshift estimates for two potential $z \sim 5$ solutions, as well as the $z \sim 17$ solution; we present the higher redshift solution in the top left panel of \fig{evs_jwst}.
This solution is in significant tension with the EVS PDF, even after accounting for stellar mass errors and the resulting Eddington bias.
We have checked the lower redshift solutions, and found that these are not in tension.

\cite{harikane_comprehensive_2022} also identify galaxies out to $z \sim 17$ in the ERO and ERS programs, covering a total area of 90 arcmin$^2$.
They use \textsc{Prospector} for photometric redshift and stellar mass estimates, and find good agreement in the stellar mass estimates for most of the objects from other studies that identified the same objects \citep{naidu_two_2022,donnan_evolution_2022,finkelstein_long_2022}.
The main exception being the $z \sim 17$ source mentioned above, for which \cite{harikane_comprehensive_2022} and \cite{donnan_evolution_2022} predict lower stellar masses (by $\sim -0.7$ dex) than those obtained by \cite{naidu_schrodingers_2022}.
The top right panel of \fig{evs_jwst} shows these candidates compared to our predicted EVS PDF; all objects lie within the contours, even at the most extreme redshifts.

Finally, in the bottom panel of \fig{evs_jwst} we show a selection of high-mass candidates from \cite{rodighiero_jwst_2022} over the 10 arcmin$^2$ area covering the SMACS0723 cluster.
They particularly target those objects that are dark in UV-optical rest-frame wavelengths, which they cite as evidence for high levels of dust obscuration.
We plot a number of their candidates, the majority of whch are consistent with our EVS predictions.
However, one candidate, at $z \sim 10$, is in significant tension.
We additionally plot the 3$\sigma$ upper limits, for a stellar fraction of unity, assuming a whole sky survey, and show that this candidate is even in tension with this highly conservative limit.

\subsection{Predictions for future surveys}

\begin{figure*}
    \includegraphics[width=0.98\textwidth]{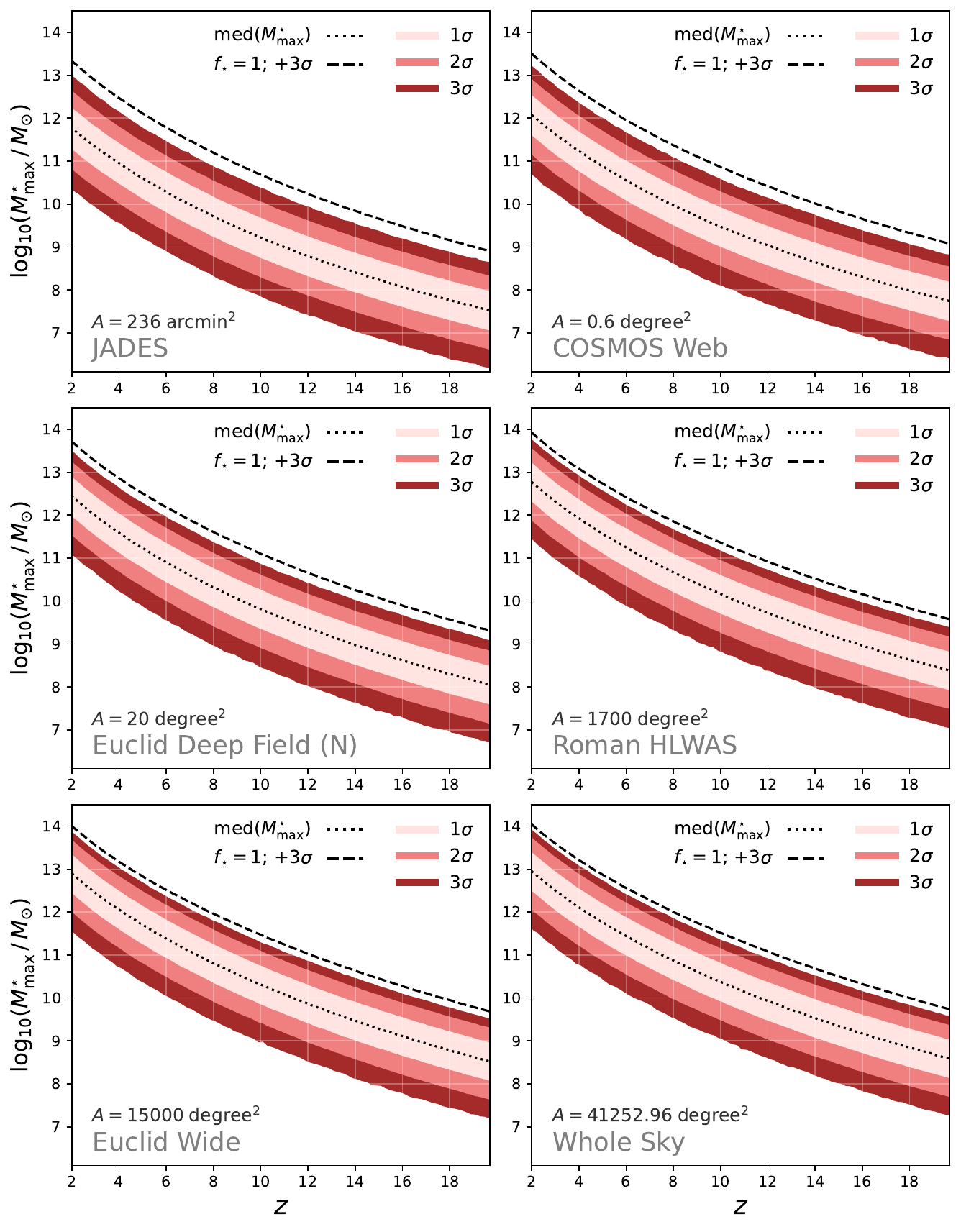}
    \caption{
    Predicted EVS PDF of the galaxy stellar mass distribution for a number of upcoming surveys.
    Clockwise from top left: the combined JADES medium and deep surveys (JWST), the COSMOS Web survey (JWST), Roman High Latitude Wide Area Survey (HLWAS), the theoretical prediction for a whole sky survey, Euclid Wide Survey, Euclid Deep Field North.
    The dashed line shows the $3\sigma$ upper limit assuming a stellar fraction of unity.
	}
    \label{fig:evs_surveys}
\end{figure*}

As well as comparing to results from existing surveys, we can also make predictions for a number of planned upcoming surveys.

There are a number of relatively wide area surveys planned with JWST
\Fig{evs_surveys} shows predictions for the full COSMOS Web survey area \citep[0.6 deg$^2$][]{kartaltepe_cosmos-webb_2021}, as well as the combined medium and deep survey areas in the JADES survey \citep[136 arcmin$^2$][]{rieke_jwst_2019}.
JWST's limited survey area is not expected to identify the rarest objects at lower redshifts, but its exceptional depth may be capable of discovering extreme objects in the epoch of very first star and galaxy formation ($z \geqslant 15$), as shown in \sec{obs_jwst}.

Wider field surveys are more likely to discover extreme objects that may challenge our cosmological and galaxy evolution models.
The Roman High Latitude Wide Area Survey, consisting of spectroscopic and imaging components \citep{wang_high_2022}, will cover an area of 1700 deg$^2$; stellar mass EVS PDF predictions for such a survey are shown in \Fig{evs_surveys}.
Euclid will also carry out a wide spectroscopic survey with the aim of constraining models of dark energy \citep[1500 deg$^{2}$]{collaboration_euclid_2022}, as well as two deep fields; we show the stellar mass EVS PDF for the planned Wide and Deep Field North (20 deg$^{2}$) surveys.

\Fig{evs_surveys} also shows the galaxy stellar mass EVS for a whole sky survey ($f_{\mathrm{sky}} = 1$).
Whole sky surveys at these high redshifts are inconceivable with current observational capabilities, however such a comparison avoids any possible \textit{a posteriori} effects of region selection.

To enable the EVS approach to be applied to arbitrary surveys, we have made a python package, \textsc{evstats}, available at \href{https://github.com/christopherlovell/evstats}{github.com/christopherlovell/evstats}, where interested users can find a simple to use Jupyter notebook detailing how to create your own confidence intervals in the stellar mass--redshift plane for a given survey area.
We also provide output files in ECSV format for the future surveys presented in \fig{evs_surveys} at \href{https://github.com/christopherlovell/evstats}{github.com/christopherlovell/evstats/tree/main/example/data}.

\section{Discussion \& Conclusions}
\label{sec:conc}

We have used extreme value statistics (EVS) to predict the stellar mass of the most massive galaxy in a flat $\Lambda$CDM universe at high redshift ($z > 5$).
Our results are as follows:
\begin{itemize}
    \item Assuming some form for the halo mass function, we calculate the EVS probability density function (PDF) for the most massive halo on a fixed-redshift hypersurface.
    The most massive halo in the \eagle\ and \flares\ simulations is within the predicted $3\sigma$ confidence intervals.
    \item We model the stellar fraction as a log-normal distribution, and combine with a fixed baryon fraction to translate our halo mass EVS PDF into one for stellar mass.
    The most massive galaxy in the \eagle\ and \flares\ simulations is within the predicted $2\sigma$ confidence intervals.
    \item We calculate the stellar mass EVS PDF for an observational survey volume, and compare to recent pre-JWST observations of galaxies at $z > 8$.
    We find tension with predicted stellar masses for two objects from \cite{tacchella_stellar_2022}, though no tension with results from \cite{harikane_search_2022}, mostly due to the significant uncertainties in the stellar mass estimates, which translate into a large Eddington bias correction.
    \item We also compare to recent high redshift candidates from the first JWST data, and find significant tension with certain stellar mass estimates of a $z \sim17$ candidate from \cite{donnan_evolution_2022, harikane_comprehensive_2022, naidu_schrodingers_2022}, as well as $z \sim 10$ candidates presented in \cite{labbe_very_2022} and \cite{rodighiero_jwst_2022}.
    \item Finally, we present the stellar mass EVS PDF for a number of upcoming surveys from JWST, Euclid and Roman between $2 < z < 20$
\end{itemize}

The use of extreme value statistics is a powerful means of understanding the likelihood of the most massive objects in the Universe, complementing existing approaches \citep{steinhardt_impossibly_2016, behroozi_most_2018,boylan-kolchin_stress_2022}.
Already a number of objects detected in recent years (pre-JWST) are in tension with the predicted distributions \citep{tacchella_stellar_2022}, assuming even conservative limits on the conversion of baryons into stars, and a number of the first candidates from early JWST data are also in significant tension \citep{donnan_evolution_2022,naidu_schrodingers_2022,labbe_very_2022,rodighiero_jwst_2022}.
We stress, however, that it is entirely plausible that these objects are not in tension with $\Lambda$CDM, and that instead there are unaccounted for uncertainties in their redshift or stellar mass estimates.

Redshift estimates of high redshift sources are often multimodal, leaving the possibility that many high redshift candidates are, in fact, low-redshift interlopers \citep[see][]{zavala_dusty_2022} and the understanding of the necessary JWST instrument calibration is evolving \citep[][]{adams_discovery_2022}.
With regards to stellar mass, estimates are sensitive to a number of modelling assumptions during the SED fitting process, such as the assumed initial mass function (IMF) and stellar population synthesis (SPS) model.
A clear example of this is the $z \sim 17$ object analysed by \cite{donnan_evolution_2022,harikane_comprehensive_2022,naidu_schrodingers_2022}; the estimates from these different studies, using a variety of different modelling assumptions, cover almost 1 dex in stellar mass.

At the highest redshifts, it is possible that Population III star formation may contribute up to 3-4 times the number of UV photons \citep{harikane_comprehensive_2022}, boosting nebular emission in the rest-frame optical, which can bias stellar mass and SFR estimates.
These first stars are also expected to have a significantly top heavy IMF, which can complicate their interpretation using standard SPS models.
The assumed prior on the star formation history can also have a large effect on derived stellar mass estimates \citep{tacchella_jwst_2022,whitler_star_2022,tacchella_stellar_2022}.
Models suggest high redshift galaxies have rising star formation histories \citep{finlator_smoothly_2011,wilkins_first_2022}; using incompatible functional forms can lead to significant biases \citep{carnall_how_2019}.
\cite{steinhardt_templates_2022} recently highlighted the impact of using templates calibrated or derived from lower redshift conditions, leading to offsets in stellar mass estimates of high redshift sources of up to 1-1.6 dex.
\cite{mason_brightest_2022} estimates the maximal UV luminosity assuming all gas in a halo is converted into stars over a timescale that maximises the UV emission ($\sim 10 \; \mathrm{Myr}$), and found that the upper limit derived is higher than that measured in recent HST \& JWST results \citep{bouwens_new_2021,donnan_evolution_2022}.
AGN contamination can also bias both stellar mass and redshift estimates \citep[see][]{inayoshi_lower_2022}, and may be a particularly pertinent contaminant in the analysis presented here; it is in the most massive halos that the most massive central black holes are expected to reside.
However, the EVS formalism presented here does allow us to place wide priors on these processes, producing self-consistent PDFs that take into account many of these uncertainties.
As our understanding of the physics of galaxy formation at high redshift improves, these priors can be narrowed, allowing for more precise limits on the maximum halo and stellar mass at a given redshift to be made.

We do not take account of the effect on our predictions of surveys taken from multiple areas of the sky.
\cite{behroozi_most_2018} argue that such a survey approach increases the chance that a single survey will contain an outlier.
However, we note that the effect of observational errors leading to Eddington and Malmquist bias has a much larger effect on the predicted probabilities.
Including this cosmic variance effect within the EVS framework is left for future work.

With upcoming wide field surveys a number of galaxies will be detected that may potentially be in tension with predictions from $\Lambda$CDM, or require extreme conversion rates of baryons into stars.
We hope the predictions presented here, and the publicly accessible code (\href{https://github.com/christopherlovell/evstats}{github.com/christopherlovell/evstats}), will present a means of producing confidence intervals for any given survey, and allow observers to quickly evaluate the probability that a given source is in tension with a given cosmology.

\section*{Acknowledgements}
We wish to thank health and other essential workers for their tireless support over the past years.
CCL wishes to thank Peter Coles for introducing him to the concept of Extreme Value Statistics, and Peter Thomas, Giulio Fabbian and Giulia Rodighiero for helpful discussions.
CCL acknowledges support from the Royal Society under grant RGF/EA/181016.
IH acknowledges support from the European Research Council (ERC) under the European Union's Horizon 2020 research and innovation programme (Grant agreement No. 849169).
We also wish to acknowledge the following open source software packages used in the analysis: \textsc{Numpy} \citep{harris2020array}, \textsc{Scipy} \citep{2020SciPy-NMeth}, \textsc{Astropy} \citep{astropy:2013,astropy:2018} and \textsc{Matplotlib} \citep{Hunter:2007}.

We list here the roles and contributions of the authors according to the Contributor Roles Taxonomy (CRediT)\footnote{\url{https://credit.niso.org/}}.
\textbf{Christopher C. Lovell}: Conceptualization, Formal analysis, Software, Writing - Original Draft. \textbf{Ian Harrison}: Methodology, Writing - Review \& Editing. \textbf{Yuichi Harikane, Sandro Tacchella, Stephen M. Wilkins}: Writing - Review \& Editing.

\section*{Data Availability Statement}
All of the code and data used in the analysis is available at \href{https://github.com/christopherlovell/evstats}{github.com/christopherlovell/evstats}.
Details on where to obtain the stellar and halo mass values from the \eagle\ and \flares\ simulations are provided in \cite{mcalpine_eagle_2016} and \cite{lovell_first_2021}, respectively.




\bibliographystyle{mnras}
\bibliography{extremes,custom} 



\appendix

\section{Redshift binning}
\label{sec:z_bin}

When drawing the confidence intervals plotted in \textit{e.g.} \fig{hypersurface_halo_EAGLE}, we sample multiple PDFs for each redshift slice.
This is equivalent to carrying out $N$ Bernoulli trials.
There is therefore a non-zero probability of $x$ events being above some threshold, which we can work out by calculating the binomial probability,
\begin{align}
    P_{x} = {N \choose x} p^{x} q^{N-x}
\end{align}
where $p$ is the chosen probability threshold.
We have tested this for the two objects from the \cite{tacchella_stellar_2022} sample shown in the top panel of \fig{evs_finkelstein21}, and find that, for our fiducial bin spacing, the probability of exceeding the $3\sigma$ contour threshold is $P_{x} < 3 \sigma$.
These objects are therefore statistically unlikely to exceed this limit due to the number of trials, and are therefore still significant outliers.


\bsp	
\label{lastpage}
\end{document}